\documentclass[11pt,a4paper]{article}
\usepackage{jheppub}
\usepackage{textcomp}
\allowdisplaybreaks
\usepackage{epsfig}

\usepackage{amsmath, amsthm, amsfonts, amssymb, latexsym}

\usepackage[caption=false]{subfig}

\usepackage{url}
\usepackage{ifpdf}

\def\beq{\begin{equation}}
\def\eeq{\end{equation}}

\def\bequ{\begin{equation}}
\def\eequ{\end{equation}}

\title{Hawking radiation for a Proca field in D-dimensions}

\author[a]{Carlos Herdeiro,} \author[a]{Marco O. P. Sampaio,} \author[a]{Mengjie Wang} 
\affiliation[a]{Departamento de F\'\i sica da Universidade de Aveiro and I3N \\ 
Campus de Santiago, 3810-183 Aveiro, Portugal} 

\emailAdd{herdeiro@ua.pt, msampaio@ua.pt, mengjie.wang@ua.pt}

 \keywords{Black holes, Proca field, Hawking radiation, TeV gravity}

\abstract{We study the wave equation of a massive vector boson in the background of a $D$-dimensional Schwarzschild black hole. The mass term introduces a coupling between two physical degrees of freedom of the field, and we solve the resulting system of ODEs numerically, without decoupling. We show how to define decoupled transmission factors from an $\mathbf{S}$-matrix and compute them for various modes, masses and space-time dimensions. The mass term lifts the degeneracy between transverse modes, in $D=4$, and excites the longitudinal modes, in particular the $s$-wave. Moreover, it increases the contribution of waves with larger $\ell$, which can be dominant at intermediate energies. The transmission factors are then used to obtain the Hawking fluxes in this channel. Our results alert for the importance of modelling the longitudinal modes correctly, instead of treating them as decoupled scalars as in current black hole event generators; thus they can be used to improve such generators for phenomenological studies of TeV gravity scenarios.}

\begin{document}
\maketitle




\section{Introduction}
Separation of variables and decoupling of degrees of freedom are two key properties in the study of wave equations in gravitational backgrounds. Most remarkably, the gravitational perturbations of the Kerr black hole both separate and decouple, as first exhibited in the celebrated work of Teukolsky \cite{Teukolsky:1972my}, which allowed to show linear stability of the Kerr solution. Introducing charge or considering higher dimensions, for instance, seems to spoil such exceptionality; indeed  gravitational perturbations do not seem separable in either the four dimensional Kerr-Newman background \cite{Chandrasekhar:1985kt} or the higher dimensional Myers-Perry background \cite{Kunduri:2006qa}.

Failure to achieve separation of variables or to decouple individual degrees of freedom typically leads to the problem of solving sets of coupled ordinary differential equations (ODEs).\footnote{Non-separability leads to partial differential equations but these may be discretised to systems of ODEs.} In the absence of analytic strategies, a full solution of the problem can only be obtained numerically.  Surprisingly, numerical solutions for such (common) problems  seem to be essentially unexplored, in the context of Hawking radiation, in contrast to the vast literature concerning fully decoupled and separable problems. Perhaps the reason is connected to the problem of quantisation, which is usually straightforward when a complete set of decoupled modes is available. We will show, however, that coupled systems (of the type considered herein) may be treated similarly using an $\mathbf{S}$-matrix type formalism which allows decoupling in the asymptotic regions.  

In this paper we shall consider a wave equation which separates but does not decouple, in the background of a $D$-dimensional spherically symmetric Schwarzschild black hole \cite{Tangherlini:1963bw}: the massive spin 1, or Proca, field. In contrast, observe that the Maxwell field in the $D=4$ Schwarzschild background has two physical decoupled polarisations. Introducing the mass term, the spin 1 field gains a longitudinal polarisation, and two of the physical degrees of freedom cannot be decoupled. In higher dimensions the situation is similar, when set in appropriate variables. We shall solve the wave equation for the Proca field in the $D$-dimensional Schwarzschild background numerically to obtain a scattering matrix. This will allow us to define transmission factors as well as the Hawking radiation flux of vector bosons from such black holes. 

The technique used herein can in principle be applied to other problems of test field wave equations or studies of gravitational perturbations, including the computation of quasi-normal modes\footnote{See \cite{Ferrari:2007rc,Molina:2010fb} for studies of quasi-normal modes where similar coupled systems have been considered.}. Our choice of the Proca field was primarily due to its simplicity, but most importantly for its phenomenological interest to TeV scale gravity scenarios \cite{Antoniadis:1990ew,Arkani-Hamed:1998rs,Antoniadis:1998ig,Arkani-Hamed:1998nn}. In such scenarios, scattering processes with centre of mass energy well above the fundamental Planck scale, should be dominated by classical gravitational interactions~\cite{'tHooft:1987rb}. Moreover, for sufficiently small impact parameter black holes should form in particle collisions, as made manifest by numerical evidence~\cite{Choptuik:2009ww}. This was predicted by early trapped surface calculations~\cite{Yoshino:2005hi} in $D\geq 4$, which gives bounds on the gravitational radiation emitted; the latter was also estimated through perturbation theory studies of head on collisions of shock waves in $D=4$. In $D>4$, recent results have produced improved semi-analytic estimates within perturbation theory~\cite{Herdeiro:2011ck}. Other estimates may be obtained from high energy black hole collisions in numerical relativity. The first results for numerical collisions of higher dimensional black holes have been  released in \cite{Witek:2010xi,Witek:2010az,Okawa:2011fv}, based on the frameworks in~\cite{Zilhao:2010sr,Yoshino:2009xp}. These black holes would then decay via Hawking radiation \cite{Hawking:1974sw} and leave observable signatures~\cite{Aad:2009wy}, such as large multiplicity of jets or large transverse momentum, and considerable missing energy due to gravitational wave emission. To better understand and model these scenarios, as well as to improve the phenomenology of black hole event generators such as \textsc{charybdis2} \cite{Frost:2009cf} and \textsc{blackmax} \cite{Dai:2007ki}, currently in use at the Large Hadron Collider, it is important to understand the Hawking radiation from higher dimensional black holes. This is especially relevant because the first experimental tests of trans-Planckian gravity scenarios are under way \cite{Khachatryan:2010wx,ATLAS-CONF-2011-065,ATLAS-CONF-2011-068}, and therefore improving their phenomenology is very timely. The computation of Hawking radiation for such models requires solving the wave equation for various fields with different spins, masses and charges in these black hole backgrounds. In particular, the Proca field in the $D$-dimensional Schwarzschild background has not been considered up to now.

Our results exhibit distinctive features as we introduce the mass term, such as the lifting of the degeneracy of the two transverse modes in four dimensions, the appearance of longitudinal mode contributions (absent for Maxwell's theory) and in particular the $s$-wave. One feature that appears not to have been discussed in the literature is that in four and five space-time dimensions, the transmission factor has a non-vanishing value in the limit of zero spatial momentum. We will also find the expected suppression with $M$, but perhaps the most relevant feature is to notice the increasing importance of the longitudinal modes and larger $\ell$ partial waves. Moreover, the precise contribution of the longitudinal modes deviates from simple models that have been used in the aforementioned Monte Carlo event generators to produce phenomenological data, providing our results with phenomenological interest in this context.

This paper is organised as follows. In Section 2 we explain how to decompose the Proca field equation in a generic factorizable background geometry consisting of a warped product of an $m$-dimensional space with an $n$-dimensional Einstein space, using the harmonic functions of the Einstein space. In Section 3 we specialise the equations of motion to a $D$-dimensional Schwarzschild background and present an appropriate set of independent degrees of freedom of the Proca field. We then study the asymptotic and near horizon behaviour of the coupled radial equations, which is required to impose the boundary conditions and to define the scattering matrix. In Section 4 we discuss how the scattering matrix is used to compute the transmission factor and the Hawking spectrum. In Section 5 we discuss the numerical method and results, and we conclude with a discussion in Section 6. Some technical relations are left to the Appendix.

\section{The Proca equations in Einstein symmetric spacetimes}

In this Section we present the wave equations for a Proca field which may be complex and charged under a $U(1)$ electromagnetic field. This covers the effective fields describing the $Z$ and $W$ particles in the Standard Model of particle physics (SM), the former being a neutral Proca field and the latter being an electromagnetically charged Proca field. In the next Section we will specialise to the neutral case which we want to study in detail. The Lagrangian is
\begin{equation}
\mathcal{L}= -\dfrac{1}{2}W^\dagger_{\mu\nu}W^{\mu\nu}+M^2W_\mu^\dagger W^\mu+iQW_\mu^\dagger W_\nu F^{\mu\nu} \ ,
\end{equation}
where $W_{\mu\nu}=\partial_\mu W_\nu-\partial_\nu W_\mu$ and we have included the coupling of the $W$ to the electromagnetic field strength tensor $F^{\mu\nu}$ as in the SM\footnote{In this paper we use the particle physics convention for the signature of the metric $\left(+---\ldots\right)$.}. The equations of motion for $W$ when all the background fields are fixed are
\begin{eqnarray}
\nabla_\nu W^{\mu\nu}-M^2W^\mu-iQW_\nu F^{\mu\nu}&=&0 \ .
\end{eqnarray}
For the gravitational background, in this Section, we consider Einstein symmetric spaces of the form~\cite{Kodama:2000fa}
\begin{equation}
ds^2=h_{ab}(y)dy^a dy^b-r(y)^2d\sigma_n^2 \ ,
\end{equation}
where $\sigma_n$ is an $n$-dimensional Einstein space with constant curvature $K$ and
\begin{equation}
d\sigma_n^2=\sigma_{ij}(x)dx^idx^j \; .
\end{equation}
Later we will specialise to the Schwarzschild black hole. We use indices $a,b,c,\ldots$ for the first set of coordinates, $\{y^a\}$, spanning the $m$-dimensional space with metric $h_{ab}(y)$; and indices $i,j,k,\ldots$ for the second set of coordinates, $\{x^i \}$, spanning the Einstein space. Furthermore, geometric quantities and differential operators on $\sigma_n$ are denoted with hats. This covers several interesting cases such as $2+n$-dimensional spherically symmetric black holes or a singly rotating black hole in $4+n$-dimensions (Kerr, Myers-Perry, etc \ldots).

To write down the equation of motion we use a decomposition of the vector field in tensorial types \cite{Kodama:2000fa}. $W_a$ are $m$-scalars, with respect to $\sigma_n$, so they must obey
\begin{equation}\label{laplace1}
\left(\hat{\Delta}+\kappa_0^2\right)W_a=0 \; ,
\end{equation}
($\kappa_0^2$ is the spin-$0$ eigenvalue).
 $W_i$ is a co-vector field which can be decomposed into a scalar $\Phi$, and a transverse co-vector $W^{T}_i$, i.e.
\begin{equation}\label{decomposition}
W_i=\hat{D}_i \Phi +W^{T}_i \;, \hspace{1cm} \; \hat{D}_i\hat{W}^{T i}=0 \ ,
\end{equation}
where $\hat{D}_i$ is the covariant derivative on $\sigma_n$ and we use an explicit hat to denote raising of indices using $\sigma_{ij}$, i.e. note that $\hat{W}^{Ti}=-r(y)^2W^{Ti}$. Since $\Phi$ is a scalar, it obeys~\eqref{laplace1}. The transverse vector obeys
\begin{eqnarray}\
\left(\hat{\Delta}+\kappa_1^2\right)W^T_i&=&0 \  , \label{laplace2}
\end{eqnarray}
where now $\kappa_1^2$ is the spin-$1$ eigenvalue.
This decomposition allows for an expansion of the various degrees of freedom $W_a,\Phi,W^T_i$ in a basis of harmonics of the Einstein space.
Furthermore, this decomposition allows for a decoupling of the field equations into an independent vector mode $W^T_i$ and $m+1$ coupled scalar fields for each set of quantum numbers labeling the basis of harmonic functions. Observe, however, that not all these modes correspond to physically independent degrees of freedom, as shown below.


\section{Neutral Proca field in a spherically symmetric black hole}\label{sec:q0Proca}
Let us now specialise to the $Q=0$ case. We shall consider separately two cases according to the value of $\kappa_0$.

\subsection{Modes with $\kappa_0\neq 0$}
Expanding the field equations with the decomposition~\eqref{decomposition}, using conditions~\eqref{laplace1} and~\eqref{laplace2}, and defining $B_a\equiv W_a-\partial_a\Phi$ we obtain
\begin{eqnarray}
\dfrac{\kappa_0^2}{r^2}B_a-\dfrac{h_{af}}{r^n}\partial_b\left[h^{db}h^{cf}r^n\left(\partial_cB_d-\partial_dB_c\right)\right]+M^2B_a+\partial_a\left[\dfrac{1}{r^{n-2}}\partial_b\left(r^{n-2}h^{bc}B_c\right)\right]&=&0 \ , \hspace{12mm}\label{Ba}\\
\dfrac{1}{r^{n-2}}\partial_a\left(r^{n-2}h^{ab}B_b\right)-M^2\Phi&=&0 \ , \\
\left[\dfrac{1}{r^2}\left(\kappa_1^2+\dfrac{\hat{R}}{n}\right)+M^2+\dfrac{1}{r^{n-2}}\partial_a\left(r^{n-2}h^{ab}\partial_b\right)\right]\hat
W^{Tj}&=&0 \label{WT}\; .
\end{eqnarray}
We consider the spherically symmetric case with $\{y^a\}=\{t,r\}$,
$|h|=1$, $h_{ab}$ is diagonal,
$h_{tt}=-1/h_{rr}\equiv V=1-\mu/r^{n-1}$ and $\mu=r_H^{n-1}$. We choose units such that the horizon radius is $r_H=1$. Since $\Phi$ is given by the second equation in terms of the other fields, it is a non-dynamical degree of freedom. In four dimensions, this agrees with the fact that a spin-1 massive field has three possible physical polarizations which in this case will be the two dynamical scalars and the transverse vector. In higher dimensions, the transverse vector on the $n$-sphere will contain more (degenerate) polarizations.

We can factor out the spherical harmonics through the expansion
\begin{eqnarray}\label{Lexpansion}
B_a&=&\beta_a^\Lambda(y)\mathcal{Y}_\Lambda(x)\ , \nonumber \\
\hat{W}^{T i}&=&q^\Lambda(y)\mathcal{Y}^i_{\Lambda}(x) \; ,
\end{eqnarray}
where $\Lambda$ denotes the mode eigenvalues for the corresponding harmonic functions. 
Furthermore, making the ansatz 
\[
\beta^{\Lambda}_t=e^{-i\omega t}\psi(r) \ , \qquad \beta^{\Lambda}_r=e^{-i\omega t}\dfrac{\chi(r)}{V} \ , \qquad q^{\Lambda}=e^{-i\omega t}\Upsilon(r)\ , 
\]
 and using~\eqref{Ba} and~\eqref{WT} we find
\begin{align}
\left[V^2\dfrac{d}{dr}\left(\dfrac{1}{r^{n-2}}\dfrac{d}{dr}r^{n-2}\right)+\omega^2-\left(\dfrac{\kappa_0^2}{r^2}+M^2\right)V \right]\chi-i\omega V^{\prime}\psi &=0\ ,\label{originalsys1}\\
\left[\dfrac{V^2}{r^n}\dfrac{d}{dr}\left(r^n\dfrac{d}{dr}\right)+\omega^2-\left(\dfrac{\kappa_0^2}{r^2}+M^2\right)V\right]\psi+i\omega\left(\dfrac{2V}{r}-V^{\prime}\right)\chi &=0 \ ,\label{originalsys2} \\
\left[\dfrac{V}{r^{n-2}}\dfrac{d}{dr}\left(r^{n-2}V\dfrac{d}{dr}\right)+\omega^2-\left(\dfrac{\kappa_1^2+\frac{\hat{R}}{n}}{r^2}+M^2\right)V\right] \Upsilon&=0 \label{transverse}
\; .
\end{align}
Thus we obtain two second order coupled radial equations for $\left\{\psi,\chi\right\}$ and a decoupled equation for $\Upsilon$. Note that $\kappa_0^2=\ell(\ell+n-1)$ and $\kappa_1^2=\ell(\ell+n-1)-1$ with $\ell$ starting at zero and one respectively. The third combination is $\kappa_1^2+\frac{\hat{R}}{n}=\ell(\ell+n-1)+n-2$.

The manipulations leading to the two coupled equations above are only valid for non-zero $M$. In the exactly massless (Maxwell) theory, a similar calculation leads to a single decoupled equation for one of the scalar modes which is
\begin{equation}
\left[\dfrac{V}{r^{n-2}}\dfrac{d}{dr}\left(r^{n-2}V\dfrac{d}{dr}\right)+\omega^2-\dfrac{\kappa_0^2}{r^2}V\right] \chi=0 \; , \label{Max}
\end{equation}
whereas the other mode $\psi=i V d_r(r^{n-2}\chi)/(\omega r^{n-2})$ is non-dynamical. Here $d_r\equiv d/dr$. The transverse mode - ruled by equation (\ref{transverse}) - remains the same for any $M$; in particular, for $M=0$, and (only) $n=2$ it becomes equivalent to (\ref{Max}). This will be manifest in the numerical results.

\subsection{Modes with $\kappa_0= 0$}
For the exceptional modes with $\kappa_0=0$, $\Phi$ does not enter the wave equation so it is a free non-dynamical field. The corresponding equation for $W^{(0)}_a$ is (the superscript denotes it is the exceptional mode)
\begin{equation}
\dfrac{h_{af}}{r^n\sqrt{|h|}}\partial_b\left[h^{db}h^{cf}r^n\sqrt{|h|}\left(\partial_cW^{(0)}_d-\partial_dW^{(0)}_c\right)\right]-M^2W^{(0)}_a=0 \; .
\end{equation}
 When $M^2\neq 0$ one uses an ansatz similar to the previous section to obtain a radial equation for a dynamical degree of freedom
\begin{equation}\label{k0M}
\left[\dfrac{V}{r^n}\dfrac{d}{dr}\left(\dfrac{r^nV}{\omega^2-M^2V}\dfrac{d}{dr}\right)+1\right]\psi^{(0)}=0 \; ,
\end{equation}
and a non-dynamical one, $\chi^{(0)}=i\omega V/(\omega^2-M^2V)d_r\psi^{(0)}$. Otherwise, for $M^2=0$, we recover the well known result that all the exceptional modes are non-dynamical (see e.g. \cite{Konoplya:2005hr}).

Now that we have covered all possibilities, several comments are in order. Firstly there is a discrete difference between the small mass limit and the exactly massless theory since we have different sets of equations for each case. This should not be surprising since there is an extra longitudinal mode for massive vector bosons. Secondly, the equations for the Maxwell theory case are all decoupled, in agreement with previous work \cite{Page:1976df}. Since decoupled radial equations have been extensively studied in the literature we will not present the details of our analysis of such modes and refer to the method in \cite{Sampaio:2009ra,Sampaio:2009tp}. Thus in the remainder we will focus on the solution of the coupled system for the massive theory, which will be used in conjunction with the decoupled modes to obtain the full Hawking spectrum in Section 5.

\subsection{Boundary conditions and radial system}\label{sec:nearhorizon}
In this section, we start by finding a series expansion of the solution near the horizon for the coupled system $\left\{\psi,\chi\right\}$. This will be used to initialise the corresponding fields for the radial integration at $r=1.001$. If we define $y=r-1$, Eqs.~\eqref{originalsys1} and
~\eqref{originalsys2} become
\begin{eqnarray}
\left[M(r)\dfrac{d^2}{dy^2}+N(r)\dfrac{d}{dy}+P(r)\right]\psi+Q(r)\chi=0\label{systerm1}\ ,\\
\left[\tilde M(r)\dfrac{d^2}{dy^2}+\tilde
N(r)\dfrac{d}{dy}+\tilde P(r)\right]\chi+\tilde Q(r)\psi=0\label{systerm2}\ ,
\end{eqnarray}
where the polynomials are defined in the Appendix. Making use of Frobenius' method to expand $\psi$ and $\chi$ as
\begin{equation}
\psi=y^\rho\sum^{\infty}_{j=0}{\mu_jy^j}\label{defpsi} \; , \qquad \chi=y^\rho\sum^{\infty}_{j=0}{\nu_jy^j} \ ,
\end{equation}
and inserting the above two equations into Eqs.~\eqref{systerm1} and
~\eqref{systerm2}, we obtain
\begin{equation}
\rho=\pm\frac{i\omega}{n-1} \; \; or \; \; \rho = 1\pm\frac{i\omega}{n-1} \; .\label{rhoexp}
\end{equation}
We want to impose an ingoing boundary condition at the horizon, so we must choose the minus sign. Furthermore, after this sign choice, the right hand side case produces a series expansion which is a special case of the left hand side (where the first coefficient is set to zero), so without loss of generality we choose $\rho = -i\omega/(n-1)$. One then writes down the recurrence relations and conclude that a general solution close to the horizon can be parametrised by the two coefficients $\nu_0$ and $\nu_1$. The other coefficients are generated by the recurrence relations~\eqref{recurone}. 

To understand the asymptotic behaviour of the waves at infinity we now study a large $r$ asymptotic expansion in the form 
\begin{equation}
\psi=e^{\beta r}r^{p}\sum_{j=0}\dfrac{a_j}{r^j}\label{psifar} \; , \qquad \chi=e^{\beta r}r^{p}\sum_{j=0}\dfrac{b_j}{r^j} \ .
\end{equation}
Inserting this into Eqs.~\eqref{originalsys1} and~\eqref{originalsys2} we obtain, at leading order,
\begin{equation}
\beta = \pm ik\; ; \qquad p = 1-\frac{n}{2}\pm i\varphi \; \; or \; \; p =-\frac{n}{2}\pm i\varphi \; ,
\end{equation}
where $\varphi=\delta_{n,2}(\omega^2+k^2)/(2k)$. Thus one can show that  asymptotically\footnote{We have used, without loss of generality, the leading power behaviour for $p$ and discarded the second option similarly to the near horizon expansion.}
\begin{equation}
\psi \rightarrow \dfrac{1}{r^{\frac{n}{2}-1}}\left[\left(a_0^++\dfrac{a_1^+}{r}+\ldots\right)e^{i\Phi}+\left(a_0^-+\dfrac{a_1^-}{r}+\ldots\right)e^{-i\Phi}\right] \ , \nonumber
\end{equation}
\begin{equation}
\chi \rightarrow \dfrac{1}{r^{\frac{n}{2}-1}}\left[\left(\left(-\frac{k}{\omega}+\dfrac{c^+}{r}\right)a_0^++\ldots\right)e^{i\Phi}+\left(\left(\frac{k}{\omega}+\dfrac{c^-}{r}\right)a_0^-+\ldots\right)e^{-i\Phi}\right] \; ,
\end{equation}
where $\Phi\equiv kr+\varphi \log r$ and $c^\pm$ is defined in the Appendix, Eq.~\eqref{eq:cpm}.
So as expected, each field is a combination of ingoing and outgoing waves at infinity. This asymptotic expansion also shows that for a generic wave at infinity, we can choose four independent quantities $\left\{a_0^\pm,a_1^\pm\right\}$ at infinity, to characterise the solution. This is expected, since we have two coupled scalar fields and for each scalar degree of freedom we must have an associated ingoing wave and outgoing wave. Thus we can define four new fields $\left\{\chi^\pm,\psi^\pm\right\}$ (which will asymptote respectively to $\left\{a_0^\pm,a_1^\pm\right\}$), by truncating the expansion for the fields and for their first derivatives at infinity. Such a transformation can be written in matrix form by defining the 4-vector $\mathbf{\Psi}^T=(\psi_{+},\psi_{-},\chi_{+},\chi_{-})$ for the new fields, and another 4-vector $\mathbf{V}^T=(\psi,d_r\psi,\chi,d_r\chi)$ for the original fields and derivatives. Then the transformation is given in terms of an $r$-dependent matrix $\mathbf{T}$ defined through
\begin{equation}
\mathbf{V}= \mathbf{T} \mathbf{\Psi} \; ,
\end{equation}
which we provide in the Appendix, Eq.~\eqref{eq:T}. Finally we obtain a first order system of ODEs for the new fields. First we define a matrix $\mathbf{X}$ through
\begin{equation}
\dfrac{d\mathbf{V}}{dr}=\mathbf{X}\mathbf{V} \; ,
\end{equation}
which is read out from the original system~\eqref{originalsys1}, \eqref{originalsys2}. Its explicit form is given in the Appendix, Eq.~\eqref{eq:X}.
Then we obtain
\begin{equation}
\dfrac{d\mathbf{\Psi}}{dr}=\mathbf{T}^{-1}\left(\mathbf{X}\mathbf{T}-\dfrac{d\mathbf{T}}{dr}\right) \mathbf{\Psi} \ .
\end{equation} We can write other equivalent systems using different $\mathbf{T}$ matrices. In particular, we have also integrated a first order system using the fields $\psi_{s}=k \psi-isd_r\psi$ and $\chi_{s}=k \chi-isd_r\chi$ which produced numerically equivalent results. The only difference is that for such fields we need to extract $\mathcal{O}(r^{-1})$ coefficients to obtain $a_1^s$. 

\section{The Hawking spectrum}
The boundary conditions we have chosen in Section~\ref{sec:nearhorizon}, are suitable for the computation of the Hawking spectrum of radiated quanta from the black hole. The Hawking spectrum is generically given by a sum over a complete set of modes with labels $\zeta$, of the transmission factor $\mathbb{T}_\zeta$ times a thermal average number of quanta produced at the horizon $\left<n_\zeta\right>$. This is defined for a basis of decoupled modes. In our problem, we have a sub-set of modes, the transverse vector mode, and the $\ell=0$ ($\kappa_0=0$) mode, which are decoupled. But we also have a tower of modes which are coupled two by two for each $\ell>0$, the two scalars $\psi$ and $\chi$. It is not obvious how to decouple them for all $r$ through an explicit transformation. Instead, let us try to understand how to extract the relevant information in the asymptotic regions.

Let us denote the two coupled fields by a 2-vector $\mathbf{U^T}=(\psi,\chi)$ and represent the coupled system of radial equations through a (linear) second order matrix differential operator $\mathcal{D}^{(2)}$ acting on $\mathbf{U}$, i.e. $\mathcal{D}^{(2)}\mathbf{U}=0$. The system is coupled because of the off diagonal elements of the $\mathcal{D}^{(2)}$ operator. To decouple the system we would have to find a transformation of the fields $\mathbf{U}=\mathcal{A} \mathbf{\bar U}$, such that the new differential operator $\bar{\mathcal{D}}^{(2)}=\mathcal{D}^{(2)}$\textopenbullet$\mathcal{A}$ is diagonal, i.e.
\begin{equation}
\bar{\mathcal{D}}^{(2)}=\left(\begin{array}{cc}
\bar{\mathcal{D}}^{(2)}_1 & 0 \\ 0 & \bar{\mathcal{D}}^{(2)}_2
\end{array}\right) \ .
\end{equation}
Even without finding such a transformation explicitly, one can draw some conclusions by assuming its existence\footnote{In fact, for example, if we consider $\mathcal{A}$ to be a general $r$-dependent matrix, we can write down two conditions for the four arbitrary functions of such a matrix. Thus, in principle, there is enough freedom.}. In particular we may establish a map between our general solution of the coupled system and the actual decoupled solution, for each of the asymptotic regions (horizon and far field). To find such a map let us first summarise the information we have on the general solution of the coupled system. 

In Section~\ref{sec:q0Proca} we have found that a general solution is parametrised by 4 independent coefficients in one of the asymptotic regions; either at the horizon or at infinity. Once we have chosen one set of coefficients, say at the horizon, due to the linearity of the equations, the 4 independent wave components at infinity are a linear combination of the 4 coefficients at the horizon. Let us formally denote the ingoing and outgoing wave coefficients at the horizon ($+/-$ respectively) by
 \[\vec{\mathbf{h}}=({\mathbf h}^+,{\mathbf h}^-)=(h^+_{i},h^-_{i}) \ , \] where $i=1,2$ since we have two fields.
Similarly, the coefficients at infinity are defined as the large $r$ limit of the $\mathbf{\Psi}$ field components (up to linear transformation which we will define next), i.e.
  \[\vec{\mathbf{y}}=({\mathbf y}^+,{\mathbf y}^-)=(y^+_{i},y^-_{i}) \ , \]
with $i=1,2$ for $\psi$ and $\chi$ respectively. Due to linearity, we can define a scattering matrix
\begin{equation}
\vec{\mathbf{y}}=\mathbf{S} \vec{\mathbf{h}} \ \ \Leftrightarrow \ \ \left(\begin{array}{c} {\mathbf y}^+ \\ {\mathbf y}^- \end{array} \right)=\left(\begin{array}{c|c} {\mathbf S}^{++} & {\mathbf S}^{+-} \\ \hline {\mathbf S}^{-+}  & {\mathbf S}^{--} \end{array} \right)\left(\begin{array}{c} {\mathbf h}^+ \\ {\mathbf h}^- \end{array} \right) \ \ \Leftrightarrow \ \ \left(\begin{array}{c} {y}^+_i \\ { y}^-_i \end{array} \right)=\sum_{j}\left(\begin{array}{c|c} S_{i j}^{++} & S_{i j}^{+-} \\ \hline S_{i j}^{-+}  & S_{i j}^{--} \end{array} \right)\left(\begin{array}{c} h_{j}^+ \\ h_{j}^- \end{array} \right) \; ,
\end{equation}
which is a set of numbers (depending on energy, angular momentum, etc\ldots) containing all the information on the scattering process. It can be fully determined by considering specific modes at the horizon and integrating them outwards. 
In our problem, we have imposed an ingoing boundary condition at the horizon which is simply ${\mathbf h}^{+}=0$. Then
\begin{equation}\label{scattering_ingoing}
{\mathbf y}^s={\mathbf S}^{s-}{\mathbf h}^{-} \; .
\end{equation}
Taking the $s=-$ component, and denoting the inverse matrix of ${\mathbf S}^{--}$ by $({\mathbf S}^{--})^{-1}$, we invert~\eqref{scattering_ingoing} to obtain the wave at the horizon given the ingoing wave at infinity
\begin{equation}
{\mathbf h}^{-} =\left({\mathbf S}^{--}\right)^{-1}{\mathbf y}^{-}\; .
\end{equation}
Inserting this relation back in the $s=+$ component of~\eqref{scattering_ingoing}, we obtain the outgoing wave in terms of the ingoing wave, at infinity
\begin{equation}\label{Reflection}
{\mathbf y}^+={\mathbf S}^{+-}({\mathbf S}^{--})^{-1}{\mathbf y}^-\equiv{\mathbf R} \, {\mathbf y}^-\; ,
\end{equation}
where in the last line we have defined the reflection matrix $\mathbf{R}$. Before proceeding, we note that there is still some freedom in the definition of the asymptotic coefficients since any (non-singular) linear combination is equally good from the point of view of satisfying the boundary condition. This freedom can be written in terms of 3 matrices $\mathbf{M}^s$, $\mathbf{M}_H^-$ relating some new fields (hatted) to the old fields
\begin{equation}
{\mathbf y}^s=\mathbf{M}^s\hat{\mathbf y}^s \; ,\; \; \; \; \; \; \; \; {\mathbf h}^{-}=\mathbf{M}_H^-\hat{\mathbf h}^{-} \; .
\end{equation} 
Since this represents the most general parametrisation of the solution in the asymptotic regions, there must be a choice which decouples the fields in those regions. To find the correct transformation we need a physical prescription. 

To obtain the transmission factor for the  decoupled components, it is instructive to remind ourselves of the calculation of the transmission factor for a single decoupled field. It is defined as the fraction of the incident wave which is transmitted to the horizon. If we look at a wave with energy $\omega$ (for an observer at infinity), with ingoing/outgoing amplitudes $Y^{(\infty)}_{\mp}$, then~\cite{Kanti:2004nr}
\begin{equation}
\mathbb{T}=\dfrac{|Y^{(\infty)}_{-}|^2-|Y^{(\infty)}_{+}|^2}{|Y^{(\infty)}_{-}|^2}=\dfrac{\omega\left(|Y^{(\infty)}_{-}|^2-|Y^{(\infty)}_{+}|^2\right)}{\omega |Y^{(\infty)}_{-}|^2}=\dfrac{\mathcal{F}^{in}_H}{\mathcal{F}^{in}_\infty} \ ,
\end{equation}
where in the last step we note that $\mathbb{T}$ can be re-expressed as a ratio between the total incident energy flux $\mathcal{F}^{in}_H$ (which is the difference between the energy carried by the ingoing wave and the energy of the outgoing wave) and the incident energy flux associated with the ingoing wave at infinity ($\mathcal{F}^{in}_\infty$). The former is the flux of energy transmitted down to the horizon.

We now compute the energy fluxes through a sphere at radius $r$ using the energy momentum tensor. This will allow us to identify the decoupled fields at infinity and at the horizon, and in particular, the ingoing and outgoing decoupled waves at infinity. Such a flux is shown to be conserved in our background, by using the conservation law for the energy momentum tensor, combined with the fact that the spatial integral of $T_t^{\phantom{t} t}$ for each energy eigen-mode is constant. It is defined, evaluated at $r$, as
\begin{equation}\label{eq:FluxR}
\mathcal{F}|_r=-\int_{S^n} d\Sigma\, T_t^{\; r}
\end{equation}
where $d\Sigma$ is the volume element on a $t,r={\rm constant}$ hyper-surface; up to an irrelevant normalisation, the energy momentum tensor for the complex neutral Proca field is
\begin{equation}
T^{\mu\nu}=-\dfrac{1}{2}\left(W^{\dagger \mu \alpha}W^{\nu}_{\; \alpha}-M^2W^{\dagger \mu}W^{\nu}+c.c.\right)-\dfrac{g^{\mu \nu}}{2}\mathcal{L} \; .
\end{equation} 
If we insert this in~\eqref{eq:FluxR}, assume a field configuration with a well defined energy $\omega$, and make use of the equations of motion, then, for the non-trivial case of $M^2\neq 0 \neq \kappa_0^2$, we obtain 
\begin{equation}\label{eq:FluxR2}
\mathcal{F}|_r=\sum_\Lambda\dfrac{i\omega V\Upsilon^{\dagger}_\Lambda}{2r^2}\dfrac{d\Upsilon_\Lambda}{dr}+\sum_\Lambda\left\{\dfrac{\kappa_0^2}{2r^2}\psi^\dagger \chi-\dfrac{1}{2M^2}\left[\dfrac{V}{r^n}\dfrac{d(r^n\xi^\dagger)}{dr}-\dfrac{\kappa_0^2\psi^\dagger}{r^2}\right]\left[i\omega\xi+\dfrac{\kappa_0^2\chi}{r^2}\right]\right\}+c.c
\end{equation}
where for convenience we define $\xi=d_r\psi+i\omega\chi$. Modes with different angular momentum eigenvalues are clearly decoupled, as are the transverse vector mode contributions in the first sum. The terms in the second sum couple two fields for fixed $\Lambda$. We can compute the flux at infinity and close to the horizon and express it in terms of the asymptotic coefficients in the corresponding region. Focusing on a specific mode and in the coupled part of the flux (second sum in~\eqref{eq:FluxR2})
\begin{equation}\label{eq:FluxInf}
\mathcal{F}^{\mathrm{coupled}}_\infty=|y_0^-|^2-|y_0^+|^2+|y_1^-|^2-|y_1^+|^2\equiv(\mathbf{y}^-)^\dagger \mathbf{y}^--(\mathbf{y}^+)^\dagger \mathbf{y}^+ \; ,
\end{equation}  
where $y_i^s$ are linear combinations of the asymptotic coefficients $a_i^s$ given in the Appendix, Eq.~\eqref{eq:yplus}. This choice of $y_i^s$ is already in a form close to decoupled, since we have separated the modulus square of the incident contribution from the reflected contribution, without interference terms. This form is invariant under separate unitary transformation of $\mathbf{y}^\pm$. Using the reflection matrix we obtain
\begin{equation}\label{eq:FluxInfT}
\mathcal{F}^{\mathrm{coupled}}_\infty=(\mathbf{y}^-)^\dagger\left(\mathbf{1}-\mathbf{R}^\dagger\mathbf{R}\right) \mathbf{y}^-\equiv (\mathbf{y}^-)^\dagger\mathbf{T}\, \mathbf{y}^-  \; ,
\end{equation}  
where we have defined a (hermitian) transmission matrix $\mathbf{T}$. This can be diagonalised through a unitary transformation which is the remaining freedom we have for $\mathbf{y}^-$. In fact we can do even better, and diagonalise the reflection matrix $\mathbf{R}$ with a bi-unitary transformation using the arbitrary unitary $\mathbf{M}^{\pm}$ transformations. Then the fields are manifestly decoupled at infinity, both at the level of the reflection matrix and the transmission matrix. As a consequence, in the decoupled basis, an incident wave is reflected back in the same decoupled mode without interference with the other mode. Finally, the transmission factors are simply the eigenvalues of $\mathbf{T}$, since they are each associated with a decoupled component. 

Furthermore, one can use the conservation law for the flux, to find an alternative expression for the transmission matrix, at the horizon (this will be useful to control numerical errors). The total flux at the horizon is
\begin{equation}\label{eq:FluxH}
\mathcal{F}^{\mathrm{coupled}}_H=\left(\mathbf{h^-}\right)^\dagger\mathbf{h^-} \; ,
\end{equation}  
where the $h^-_i$ coefficients are linear combinations of the two independent $\nu_i$ coefficients ($i=0,1$), given in the Appendix, Eqs.~\eqref{eq:hminus}. Eq.~\eqref{eq:FluxH} establishes the important point that the flux is positive definite, so the transmission factors must be positive definite (as expected since there is no superradiance in Schwarzschild spacetime). Finally, using the relation between $\mathbf{y}^-$ and $\mathbf{h}^-$ through\footnote{Note that the relation between $\mathbf{h}^-$ and $(\mathbf{M}^-)^{-1}\mathbf{y}^-$ can be made diagonal using $\mathbf{M}^{-}_H$, so the problem is also decoupled at the horizon.} $\mathbf{S}^{--}$, we find
\begin{equation}\label{eq:FluxH2}
\mathbf{T}=(\mathbf{S}^{--}\mathbf{S}^{\dagger--})^{-1} \; .
\end{equation}  
Once we have obtained the transmission factors, the number and energy fluxes are given by the standard result
\begin{equation}\label{eq:HawkFlux}
\dfrac{d\left\{N,E\right\}}{dt d\omega}=\dfrac{1}{2\pi}\sum_\ell\sum_\zeta \dfrac{\left\{1,\omega \right\}}{\exp(\omega/T_H)-1}d_\zeta \mathbb{T}_{\zeta} \ ,
\end{equation}
where $\zeta$ is a label running over the final set of decoupled scalar modes and the transverse mode, and $d_\zeta$ are the degeneracies of the corresponding spherical harmonics. Labeling the scalar and vector harmonic degeneracies by $d_S$ and $d_V$ respectively we have \cite{Ishibashi:2011ws}
\begin{eqnarray}\label{eq:degen}
d_S&=& \dfrac{(n+2\ell-1)(n+\ell-2)!}{(n-1)!\ell!}\ , \\
d_V&=& \dfrac{(n+2\ell-1)(n+\ell-1)(n+\ell-3)!}{(\ell+1)(\ell-1)!(n-2)!} \; .
\end{eqnarray}
The Hawking temperature in horizon radius units is 
\begin{equation}
T_H=\dfrac{n-1}{4\pi} \; .
\end{equation}

\section{Results}
\label{results}
In this Section we present a selection of numerical results to illustrate the behaviour of the transmission factors and the corresponding Hawking fluxes. To integrate the coupled and decoupled radial equations, we wrote firstly test codes in \textsc{mathematica}7 and then a code in the \textsc{c++} language, using the numerical integration routines of the Gnu Standard Library (GSL). Besides using different programming frameworks we have also tested different integration strategies which all agreed within relative numerical errors smaller than 0.1~\%. In fact, most of our numerical points have a precision which is one order of magnitude better. To check numerical errors we have integrated the radial equations up to a large radius of typically $r=10^4r_H$ and varied this up to a factor of 3 to check the precision. Furthermore we have used the two expressions for the transmission factor from Eq.~\eqref{eq:FluxInfT} and~\eqref{eq:FluxH2} which agree within the quoted precision for almost all energies. The exception is for small energy, where the first definition converges poorly. This can be explained by a simple analysis of propagation of errors combined with the fact that the $\mathbf{y}^{\pm}$ coefficients grow very fast as we decrease energy, thus requiring a very large precision for some fine cancellations to occur. The second expression is thus more natural in that limit since it does not need such cancellations and does not require such large precision.

\begin{figure}[t]
\includegraphics[scale=0.68,clip=true,trim= 0 0 0 0]{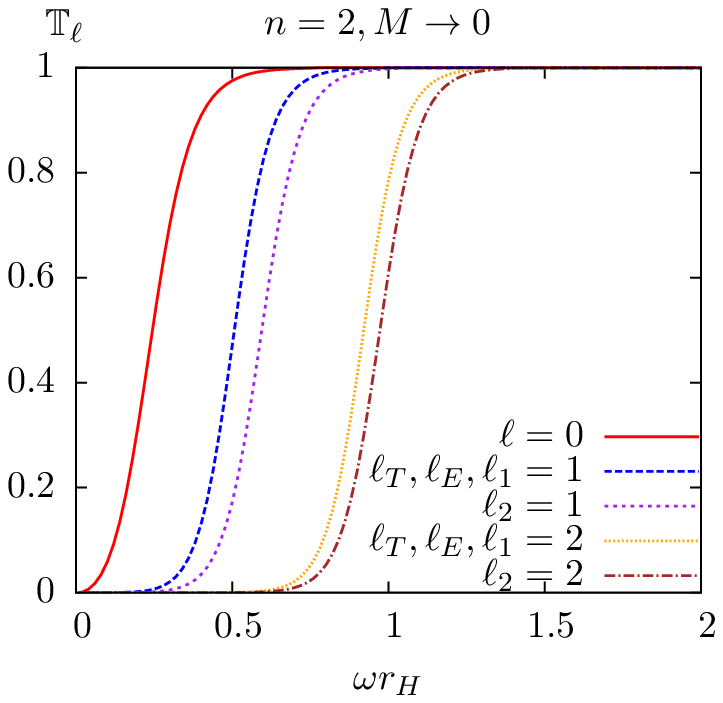}\hspace{0mm}  \includegraphics[scale=0.68,clip=true,trim= 0 0 0 0]{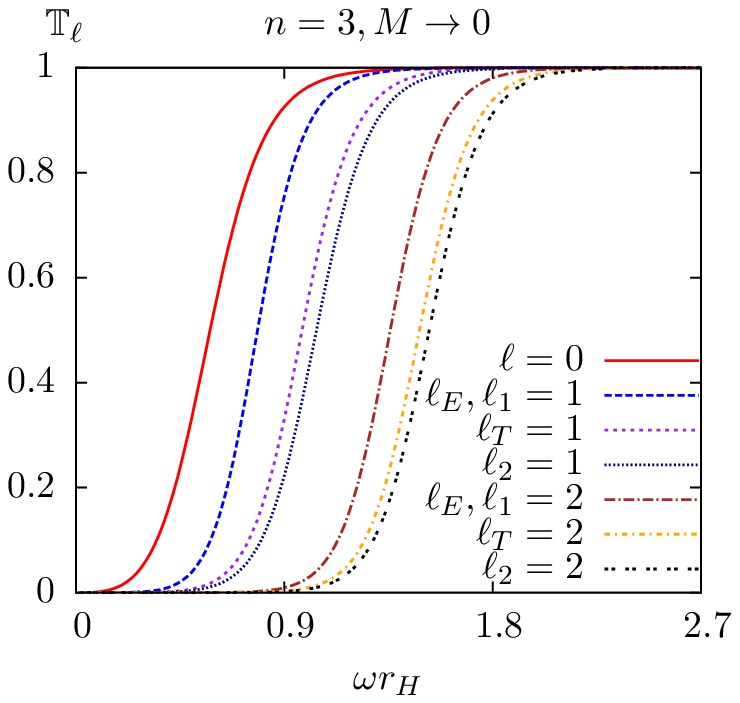} \hspace{-2.5mm} \includegraphics[scale=0.68,clip=true,trim= 0 0 0 0]{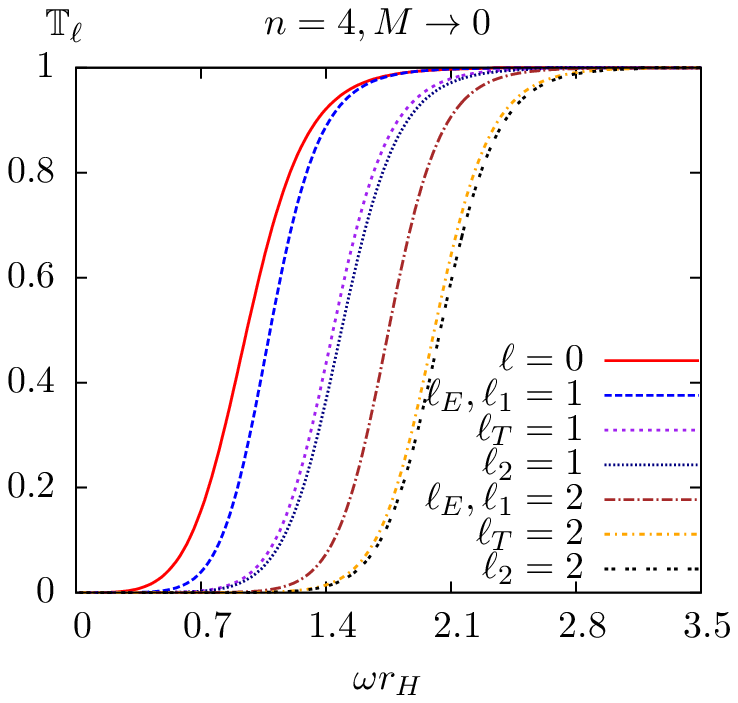} \vspace{0mm}\\
\includegraphics[scale=0.68,clip=true,trim= 0 0 0 0]{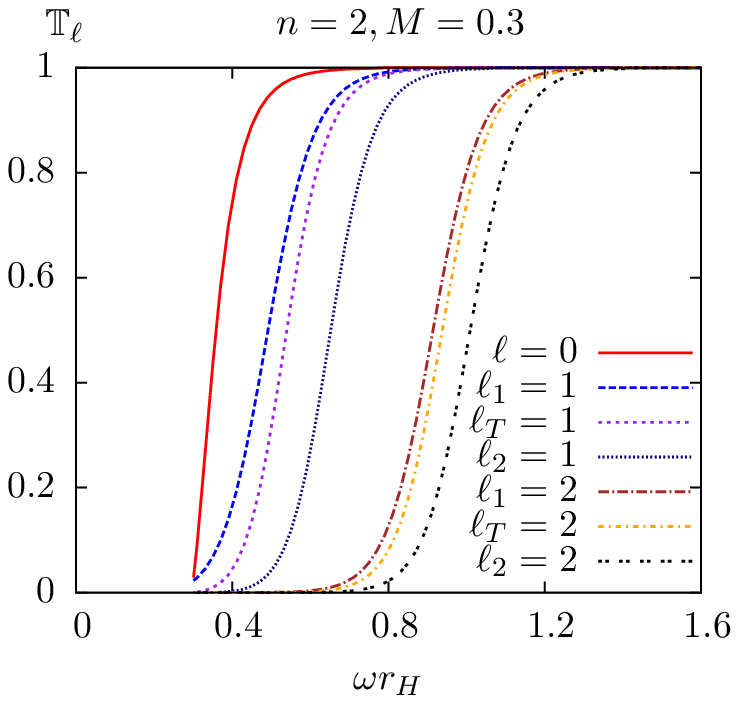} \hspace{-2.5mm} \includegraphics[scale=0.68,clip=true,trim= 0 0 0 0]{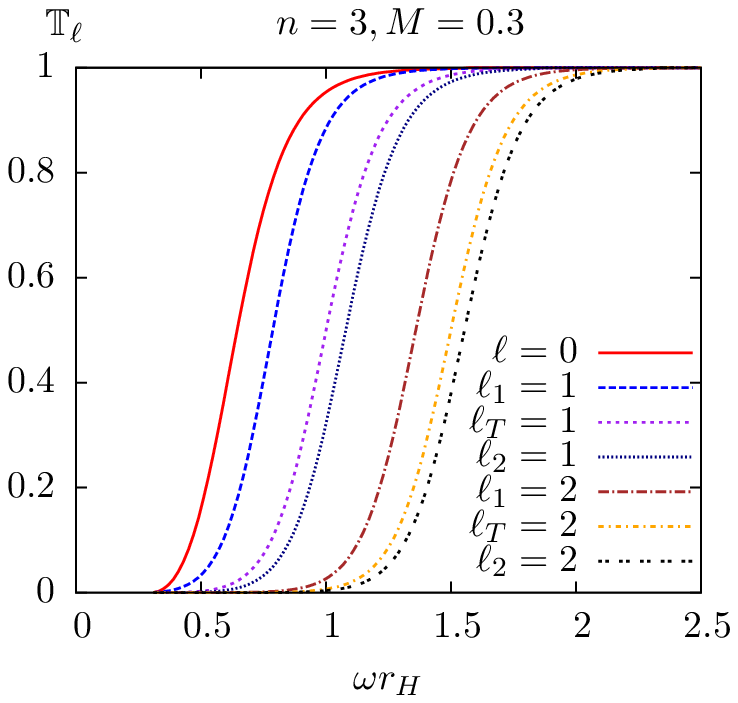} \hspace{-2.5mm} \includegraphics[scale=0.68,clip=true,trim= 0 0 0 0]{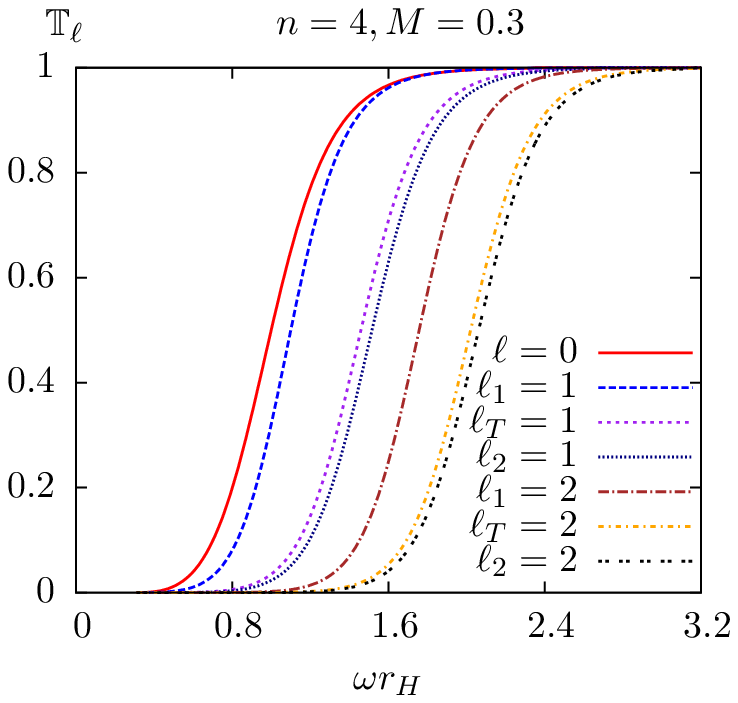} \vspace{0mm} \\
\includegraphics[scale=0.68,clip=true,trim= 0 0 0 0]{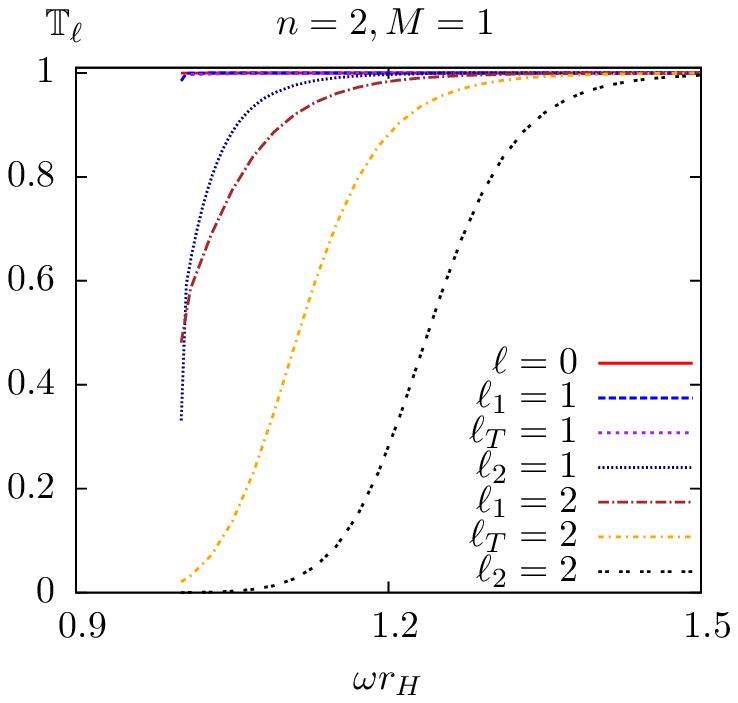} \hspace{-2.5mm} \includegraphics[scale=0.68,clip=true,trim= 0 0 0 0]{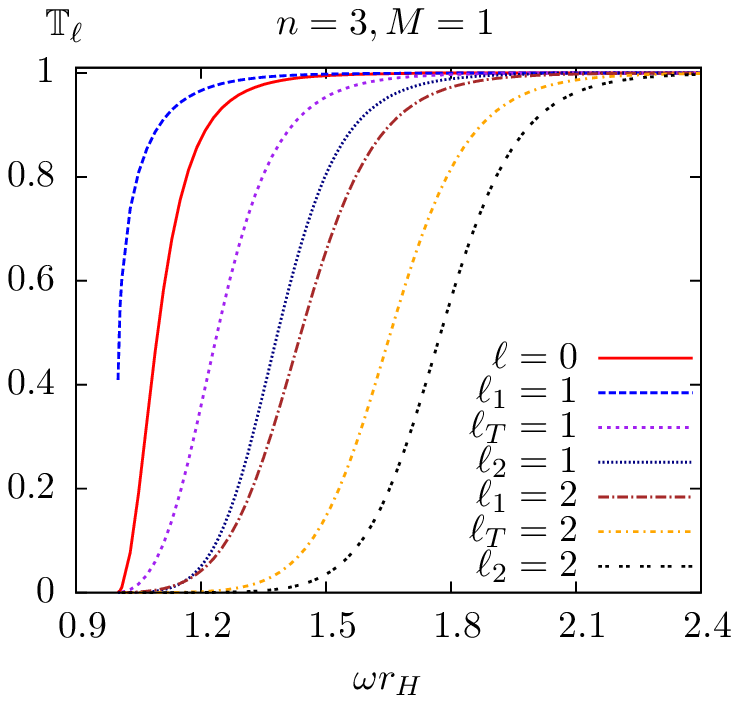} \hspace{-2.5mm} \includegraphics[scale=0.68,clip=true,trim= 0 0 0 0]{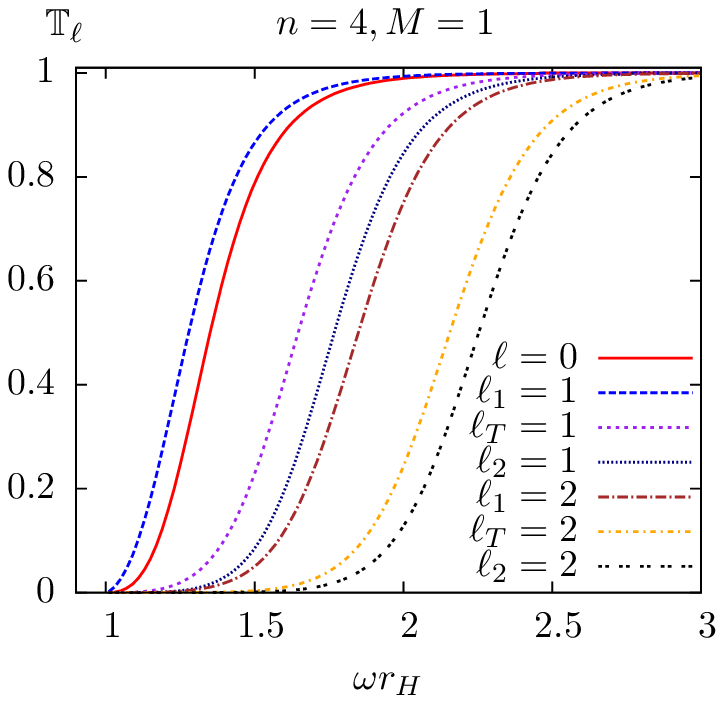}
\caption{\label{fig:Tfacs} {\em Transmission factors:} The three rows of panels, show the first few partial waves contributing to the Hawking spectrum. Each row corresponds to a fixed mass and each column to a fixed dimension. In particular, the first row shows the small mass limit of the Proca theory in order to compare it with Maxwell's theory.
}
\end{figure}
We have generated several samples of transmission factors, some of which are displayed in Fig.~\ref{fig:Tfacs}. Hereafter, we shall denominate the partial waves associated to the different modes of the Proca field by $\ell_1,\ell_2,\ell_T$ and $\ell=0$, where $\ell_1,\ell_2$ correspond to the two coupled modes ruled by Eqs. (\ref{originalsys1}) and  (\ref{originalsys2}), $\ell_T$ to the decoupled mode described by  Eq.~\eqref{transverse} and $\ell=0$ to the $\kappa_0=0$ mode, ruled by Eq.~\eqref{k0M}. Moreover, partial waves associated to the Maxwell field shall be denoted by $\ell_E$, and are ruled by~\eqref{Max}.

In the top row panels of Fig.~\ref{fig:Tfacs}, we show the partial wave contributions for $n=2,3,4$ in the zero mass limit. Some general properties are as follows. The $\mathbb{T}_\ell$ curve becomes shifted towards higher frequencies both as $\ell$ is increased, for $n$ fixed, and as $n$ is increased, for $\ell$ fixed. The former can be understood from standard geometrical optics arguments. Moreover, for this choice, there is always a numerical coincidence between one of the partial waves ($\ell_1$) obtained from the two coupled fields and the electromagnetic partial wave $\ell_E$. The $\ell=0$ and $\ell_2$ modes are always absent in the Maxwell theory, so they can be associated with the longitudinal polarization of the massive vector field. Similarly, the $\ell_T$ and $\ell_1$ partial waves are associated with the transverse polarizations of the field. A qualitative dependence on dimension is that for $n=2$, $\ell_T$ and $\ell_1$ (or $\ell_E$) modes are all equal. Curiously, this is in agreement with the fact that they describe the same number of transverse degrees of freedom as can be seen from the degeneracies~\eqref{eq:degen} specialised for $n=2$. This degeneracy is lifted for $n>2$.

For non-zero mass (middle and bottom row panels of Fig.~\ref{fig:Tfacs}), the degeneracy observed for $n=2$ in the massless limit is lifted. Also, we observe, for all $n$, that modes with higher $\ell$ partial waves  (especially $\ell_1$ modes) become a more dominant contribution at lower energies, as compared to lower $\ell$ partial waves of other modes. In particular for $M=1$, the transmission factor for $\ell_1$ becomes the largest for small energy. This effect of excitation of sub-dominant partial waves is well known to exist for example as we increase $n$ (and we can also observe such effect in our plots) as well as with the introduction of black hole rotation \cite{Marcothesis}. If this effect persists cumulatively on a rotating background, then we may have enhanced angular correlations for massive Proca fields emitted from the black hole, since higher $\ell$ partial waves are less uniform.  

Another outstanding point is that for large mass, when $n=2,3$, it can be seen that the transmission factor starts from a constant non-zero value at the threshold $\omega=M$ ($k=0$), at least for small $\ell$. We have checked that this does not happen for $n\geq 4$ for masses as large as $M=10\sim 15$, where the curves always asymptote smoothly to zero at $k=0$. Note that the parameter in the radial equations is $M^2$ so these are very large masses. A possible explanation for this phenomenon can be motivated from considerations about the range of the gravitational field in Rutherford scattering. In $n=2$, the total cross-section for Rutherford scattering diverges, so the Newtonian gravitational potential is long ranged. This means that the effective size of the gravitational potential is infinite. The same happens in $n=3$ but only at zero momentum $k=0$. This indicates that a possible reason is that an incident wave at infinity with a very small momentum will still be sufficiently attracted by the gravitational field so that a constant non-zero fraction is still absorbed by the potential. In particular we note that some of the radial equations are similar in form to those obeyed by massive scalar and massive fermion fields, so the same effect exists for such fields. To our knowledge, this feature has not been noted or discussed in the literature. The only exception is the paper by Nakamura and Sato~\cite{Nakamura:1976nc} in four dimensions, where it is claimed that the reflection factor for a scalar field always goes to $1$ at $\omega=M$ (and thus the transmission factor goes to zero). Their result seems, however, inconsistent with Fig. 1,2 and~3 of the paper by Page~\cite{Page:1977um} (also in four dimensions), where the Hawking fluxes for massive fermions become constant at the $k=0$ threshold (in agreement with our result).

\begin{figure}[t]
\includegraphics[scale=0.65,clip=true,trim= 0 0 0 0]{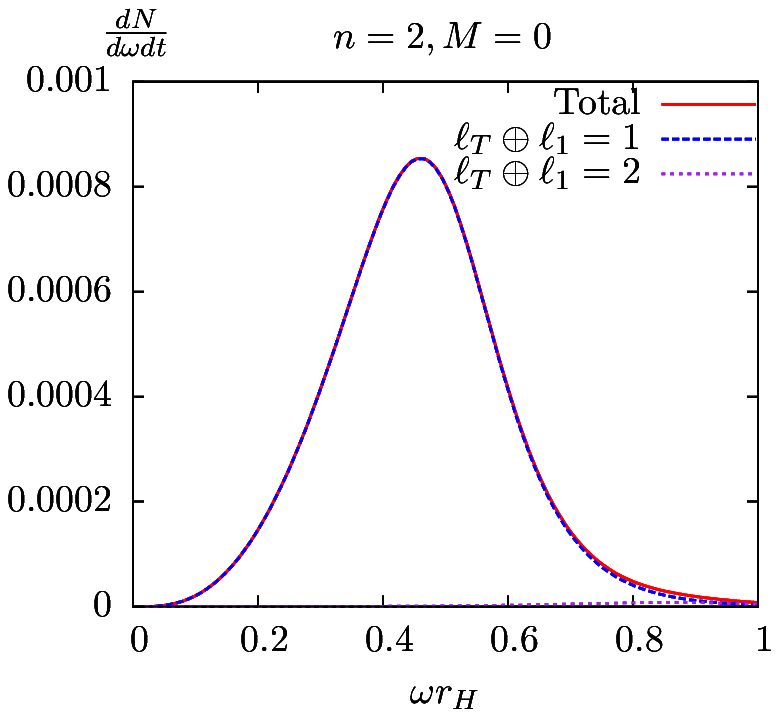}\hspace{0mm}  \includegraphics[scale=0.65,clip=true,trim= 0 0 0 0]{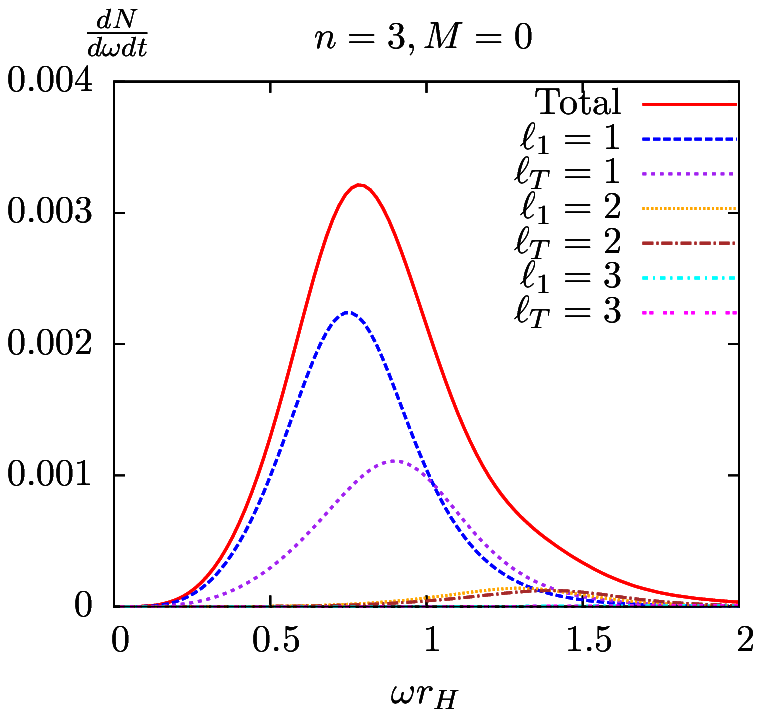} \hspace{-2.3mm} \includegraphics[scale=0.64,clip=true,trim= 0 0 0 0]{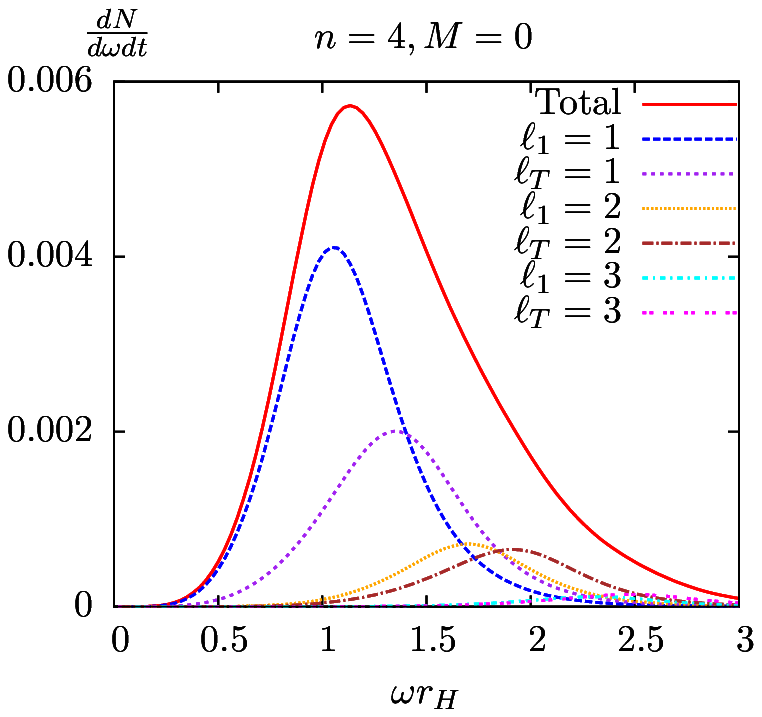} \vspace{0mm}\\
\includegraphics[scale=0.653,clip=true,trim= 0 0 0 0]{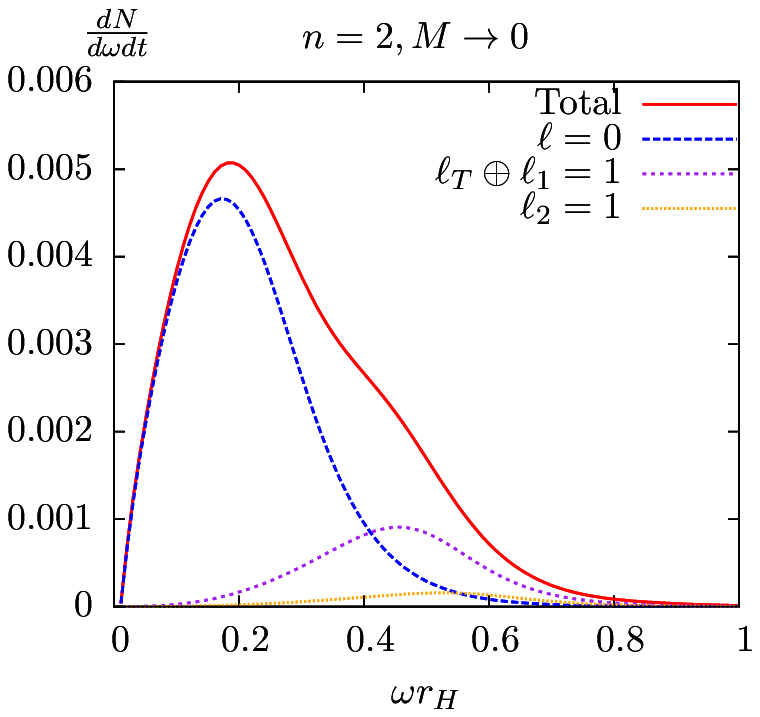} \hspace{-2.5mm} \includegraphics[scale=0.653,clip=true,trim= 0 0 0 0]{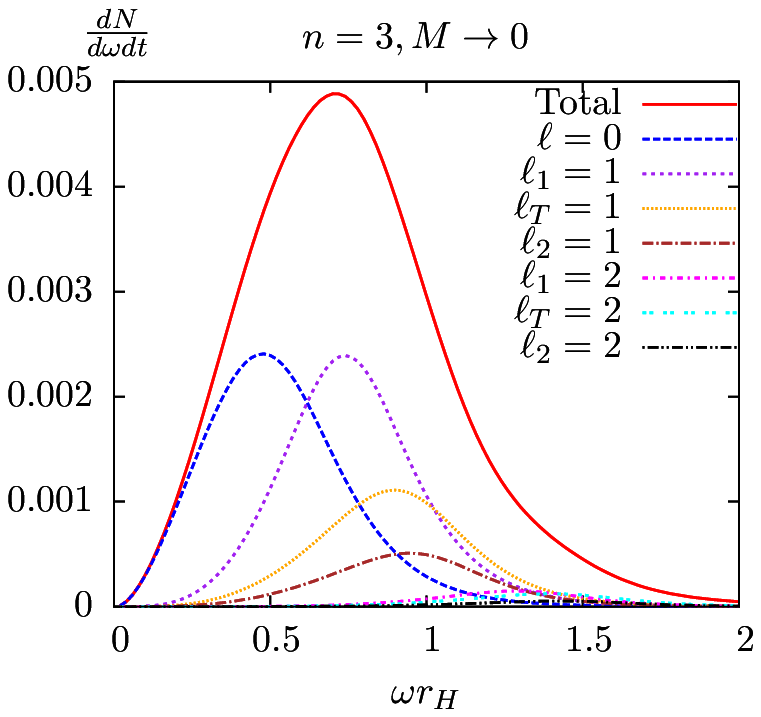} \hspace{-2.5mm} \includegraphics[scale=0.653,clip=true,trim= 0 0 0 0]{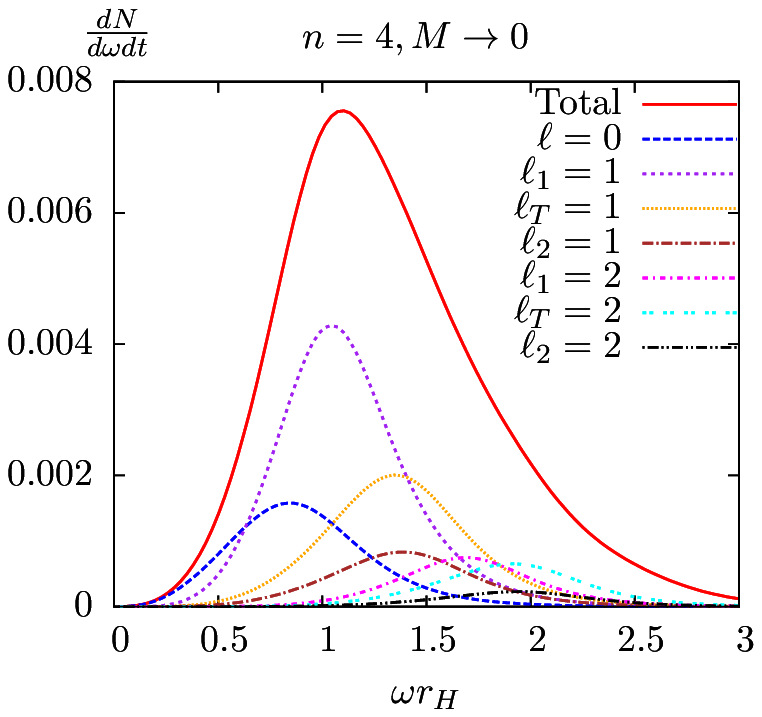} \vspace{0mm} 
\caption{\label{fig:NfluxM0} {\em Number fluxes for $M=0$ (top panels) and $M\rightarrow 0$ (bottom panels):} The red solid curve of the top panels shows the Hawking flux of particles summed over the dominant partial waves for the Maxwell theory. The different partial waves are multiplied by the corresponding degeneracies. In the bottom panels the small (but non zero) $M$ limit of the Proca theory is shown for comparison. The $\oplus$ symbol denotes the addition of modes which are numerically equal.}
\end{figure}
Once we obtain the transmission factors, the computation of the Hawking fluxes~\eqref{eq:HawkFlux} follows straightforwardly by summing up partial waves with the appropriate degeneracy factors~\eqref{eq:degen}. We have chosen to show the flux of number of particles. The flux of energy has similar features and is simply related by multiplying each point in the plots by $\omega$.

In Fig.~\ref{fig:NfluxM0} we compare the Hawking fluxes of the Maxwell theory with the small mass limit of the Proca theory. For the particular case of $n=2$ we have reproduced the results by Page~\cite{Page:1976ki} for the electromagnetic field and found very good agreement.  All panels show a red solid curve corresponding to the total Hawking flux summed up over partial waves. The partial waves included in the sum are also represented, scaled up by the appropriate degeneracy factor. As claimed in the discussion of the transmission factors, as we increase $n$, partial waves with larger $\ell$ become more important, for both Maxwell and Proca fields. One can clearly see that there is a large contribution to the total flux from the longitudinal degrees of freedom, since the vertical scales are larger for the Proca field. In particular the $\ell=0$ mode enhances the spectrum greatly at small energies. Note that these extra contributions associated with the longitudinal degrees of freedom cannot in general (for arbitrary mass) be described by a scalar field, since there is always a contribution from the coupled modes $\ell_1,\ell_2$. That is, however, the approximation done so far in black hole event generators, where the $W$ and $Z$ fields Hawking spectra in use are those of the electromagnetic field (for transverse polarizations) and a scalar field (for the longitudinal polarization). Thus, our results can be readily applied to improve this phenomenological modeling.

\begin{figure}[t]
\includegraphics[scale=0.652,clip=true,trim= 0 0 0 0]{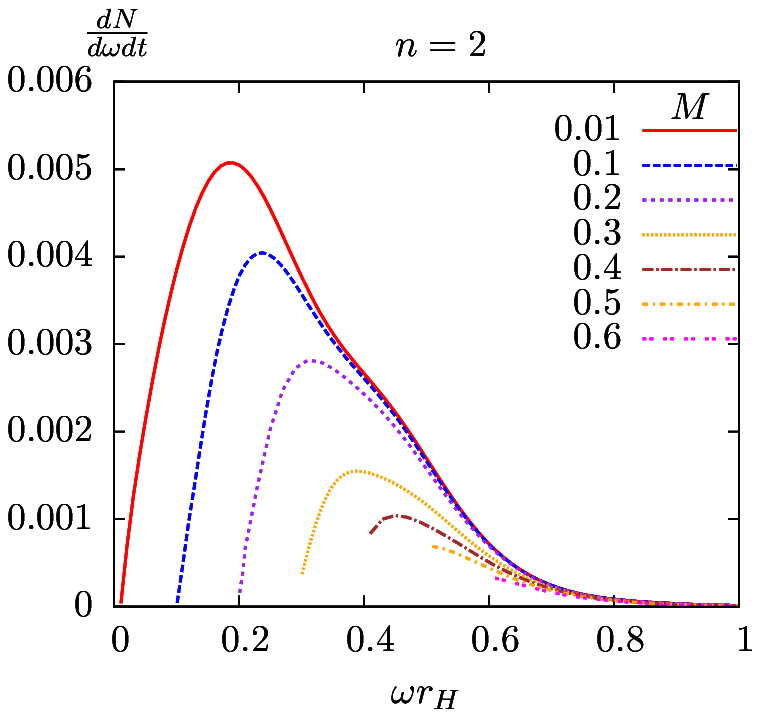} \hspace{-2.5mm} \includegraphics[scale=0.652,clip=true,trim= 0 0 0 0]{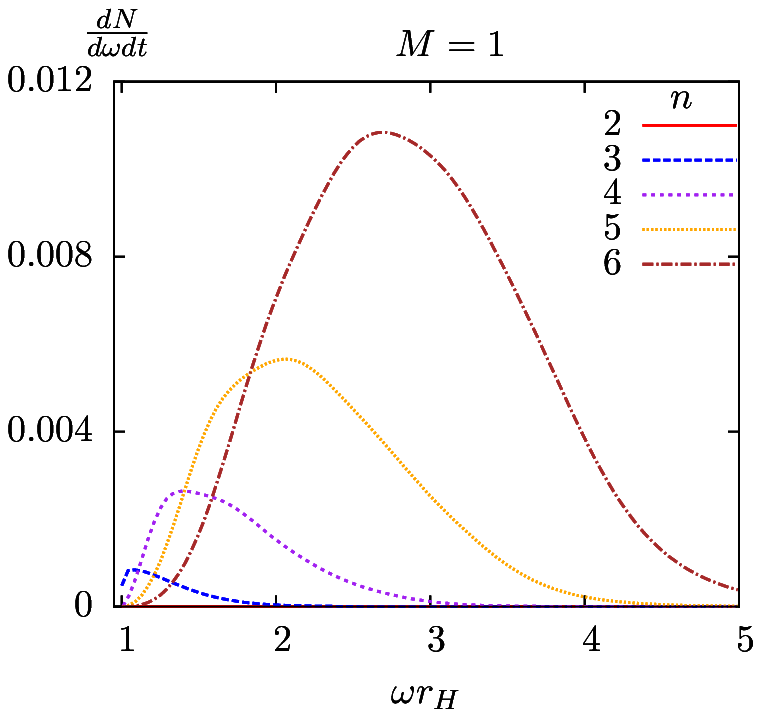} \hspace{-2.5mm} \includegraphics[scale=0.652,clip=true,trim= 0 0 0 0]{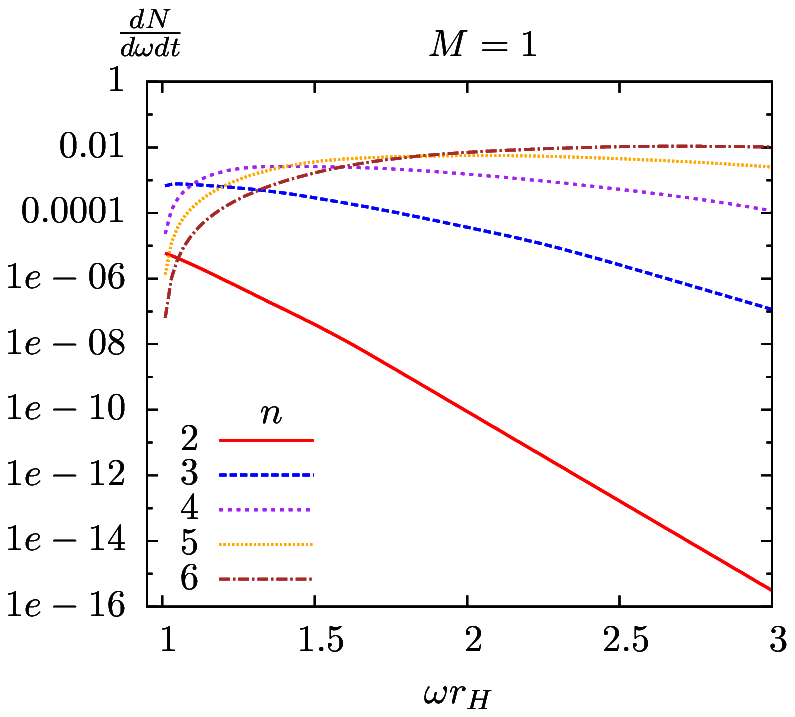} \vspace{0mm}
\caption{\label{fig:NfluxComparison} {\em Number fluxes for various $M$ and $n$:} (Left panel) Variation of the flux of particles for fixed $n=2$ and variable mass. (Middle and right panels) Variation of the flux with $n$ in a linear and logarithmic scale respectively. The logarithmic scale shows more clearly that the limiting flux at $k=0$ is finite for $n=2,3$.}
\end{figure}
In Fig.~\ref{fig:NfluxComparison} we show the variation of the total number flux with $n$ and $M$. The left panel shows the expected variation with $M$: that the flux not only gets cutoff at the energy threshold $\omega=M$, but it is also suppressed with $M$ (the same holds for $n>2$). This is the same behaviour as found in~\cite{Sampaio:2009tp,Sampaio:2009ra}. As pointed out already, in event generators massive vector particles are modeled using the Hawking fluxes for the Maxwell field and a massless scalar, with a cutoff at the mass threshold. In \cite{Sampaio:2009ra,Sampaio:2009tp} it was shown that simply imposing a sharp cut-off on the fluxes of massless scalars and fermions over-shoots the real amount of Hawking radiation emitted in the massive scalar and fermion channel. Qualitative inspection of our results suggests a similar effect for the $W$ and $Z$ channels in the evaporation. A quantitative comparison, however, requires a consideration of a Proca field confined to a thin brane. The middle and right panel show variation with $n$. In addition to the well known large scaling of the  area under the curve and the shift of the spectrum to larger energies, we can also see that more partial waves start contributing to the shape of the curve which becomes more wavy. This is particularly true because the degeneracy factors for fixed $\ell$ increase rapidly with $n$, which is a consequence of the larger number of polarizations available for a vector boson in higher dimensions. Finally, regarding $n=2,3$ we confirm the feature that the flux becomes a constant at $k=0$. This can be seen more clearly in the right panel in a logarithmic scale where the lines for $n\geq 4$ curve very rapidly around that point, whereas for $n=2,3$ they tend to a constant.

\section{Conclusion}

In this paper we have used a numerical strategy to solve the coupled wave equations obtained in the study of a Proca field in the background of a $D$-dimensional Schwarzschild black hole. Our results show some expected features, such as the mass suppression of the Hawking fluxes as the Proca mass is increased, but also some novel features, such as the non-zero limit of the transmission factor, for vanishing spatial momentum, in $n=2,3$. Moreover, for the first time a precise study of the longitudinal degrees of freedom was carried out.

One application of our results will be to improve the model in the \textsc{charybdis2} Monte Carlo event generator \cite{Frost:2009cf} currently in use by the ATLAS and CMS experiments. This simulates the production and decay of higher dimensional black holes in parton-parton collisions, a scenario to be soon tested at the LHC 14TeV centre of mass energy collisions. The Hawking evaporation can still be improved greatly through the numerical study of various wave equations in black hole backgrounds, which approximate the ones that could be produced at the LHC. This is illustrated by our results in this paper, which show the inaccuracy of the naive modeling of longitudinal modes by scalar fields, an approximation that has been used in event generators. For the phenomenology of TeV gravity scenarios, an important extension of the results herein is to consider a Proca field on the brane for scenarios where the vector bosons are confined within a thin brane. We hope to report on such results soon.

\section*{Acknowledgments}
We thank Vitor Cardoso for comments on a draft of this paper. 
MS and MW are supported by FCT-Portugal through grants SFRH/BPD/69971/2010 and a research grant from the project CERN/FP/116341/2010, respectively. This work was partly supported by the project PTDC/FIS/098962/2008.

\appendix

\section{Functions and matrices}

\begin{align}
M(r)&\equiv \sum^{2n-1}_{m=0}{\alpha_m y^m}= r\left(r^{n-1}-1\right)^2 \; ,\nonumber\\
N(r)&\equiv \sum^{2n-2}_{m=0}{\beta_m y^m}=n\left(r^{n-1}-1\right)^2  \; ,\nonumber\\
P(r)&\equiv \sum^{2n-1}_{m=0}{\gamma_m y^m}=-(\kappa_0^2+M^2r^2)\left(r^{2n-3}-r^{n-2}\right) +\omega^2r^{2n-1}  \; ,\nonumber\\
Q(r)&\equiv \sum^{2n-2}_{m=0}{\sigma_m y^m}= i\omega r^{n-1}\left(2r^{n-1}-n-1\right) \; ,\nonumber\\
\tilde M(r)&\equiv \sum^{2n}_{m=0}{\tilde \alpha_m y^m}=r^2\left(r^{n-1}-1\right)^2  \; ,\nonumber\\
\tilde N(r)&\equiv \sum^{2n-1}_{m=0}{\tilde \beta_m y^m}=(n-2)r\left(r^{n-1}-1\right)^2\nonumber,\\
\tilde P(r)&\equiv \sum^{2n}_{m=0}{\tilde \gamma_m y^m}=-\left(\kappa_0^2+M^2r^2\right)\left(r^{2n-2}-r^{n-1}\right)
+\omega^2r^{2n}-(n-2)\left(r^{n-1}-1\right)^2,\nonumber\\
\tilde Q(r)&\equiv \sum^{n}_{m=0}{\sigma_m y^m}=-i\omega(n-1)r^n\nonumber.
\end{align}
The recurrence relations are
\begin{eqnarray}
\mu_0&=&\nu_0\;,\nonumber\\
\mu_1&=&-\frac{\left(\rho(\rho-1)\alpha_3+\rho\beta_2+\gamma_1+\sigma_1\right)\nu_0+\sigma_0\nu_1}{\rho(\rho+1)\alpha_2+\gamma_0}\;,\nonumber\\
\mu_j&=&\frac{\omega^2+(n-1)^2(\rho+j)(\rho+j-1)}{D_j}f_j+\frac{i\omega(n-1)}{D_j}\tilde
f_j\;,\nonumber\\
\nu_j&=&\frac{\omega^2+(n-1)^2(\rho+j)(\rho+j-1)}{D_j}\tilde
f_j+\frac{i\omega(n-1)}{D_j}f_j\;,
\label{recurone}
\end{eqnarray}
with
\begin{eqnarray}
D_j&=&(n-1)^2\omega^2+\left(\omega^2+(n-1)^2(\rho+j)(\rho+j-1)\right)^2\;,\nonumber\\
f_j&=&-\sum^j_{m=1}{\Big[\big(\alpha_{m+2}(\rho+j-m)(\rho+j-m-1)+\beta_{m+1}(\rho+j-m)+\gamma_m\big)\mu_{j-m}+\sigma_m\nu_{j-m}\Big]}\;,\nonumber\\
\tilde
f_j&=&-\sum^j_{m=1}{\left[\big(\tilde\alpha_{m+2}(\rho+j-m)(\rho+j-m-1)+\tilde\beta_{m+1}(\rho+j-m)+\tilde\gamma_m\big)\nu_{j-m}+\tilde\sigma_m\mu_{j-m}\right]}\;.\nonumber
\end{eqnarray}
The coefficients used in the asymptotic expansion in the text are
\begin{equation}\label{eq:cpm}
c^\pm=\dfrac{i}{2\omega}\left[-\kappa_0^2+\dfrac{n(2-n)}{4}+(2\omega^2-M^2)\left(\delta_{2,n}+\delta_{3,n}\right)+\left(\pm i\dfrac{M^2}{2k}-\left(\dfrac{M^2}{2k}\right)^2\right)\delta_{2,n}\right] \; .
\end{equation}
The matrices used in the text are as follows
\begin{equation}\label{eq:T}
\mathbf{T}=\frac{1}{r^{\frac{n-2}{2}}}\left(\begin{array}{cccc}\frac{e^{i\Phi}}{r} & \frac{e^{-i\Phi}}{r} & e^{i\Phi} & e^{-i\Phi} \vspace{2mm}\\ \frac{ike^{i\Phi}}{r} & -\frac{ike^{-i\Phi}}{r} & \left[ik+\frac{i\varphi-\frac{n-2}{2}}{r}\right]e^{i\Phi} & -\left[ik+\frac{i\varphi+\frac{n-2}{2}}{r}\right]e^{-i\Phi} \vspace{2mm}\\ 0 & 0 & \left(-\frac{k}{\omega}+\frac{c^+}{r}\right)e^{i\Phi}  & \left(\frac{k}{\omega}+\frac{c^-}{r}\right)e^{-i\Phi} \vspace{2mm}\\ 0 & 0 & \left[-\frac{ik^2}{\omega}+\frac{ikc^+-\frac{k}{\omega}(i\varphi-\frac{n-2}{2})}{r}\right]e^{i\Phi}& -\left[\frac{ik^2}{\omega}+\frac{ikc^-+\frac{k}{\omega}(i\varphi+\frac{n-2}{2})}{r}\right]e^{-i\Phi} \end{array}\right) \ ,
\end{equation}
\begin{equation}\label{eq:X}
\mathbf{X}=\left(\begin{array}{cccc}0 & 1 & 0 & 0 \vspace{2mm} \\ -\dfrac{P}{M} & -\dfrac{N}{M}& -\dfrac{Q}{M}  & 0 \vspace{2mm}\\0 & 0 & 0 & 1 \vspace{2mm} \\ -\dfrac{\tilde Q}{\tilde M}  & 0& -\dfrac{\tilde P}{\tilde M}&-\dfrac{\tilde N}{\tilde M} \end{array}\right) \ ,
\end{equation}
and the 2-vectors
\begin{equation}\label{eq:yplus}
\mathbf{y}^{\pm}=\left(\begin{array}{c}\sqrt{\dfrac{\kappa_0^2 k}{\omega}} a_0^{\pm} \\ \sqrt{\dfrac{\omega k}{M^2}}\left[(\pm i\varphi-\frac{n-2}{2}+i\omega c^{\pm}\mp ik\delta_{n,2})a_0^{\pm}\pm ika_1^{\pm}\right] \end{array}\right) \ ,
\end{equation} 
\begin{equation}\label{eq:hminus}
\mathbf{h}^-=\left(\begin{array}{c}\sqrt{\kappa_0^2} \nu_0 \\ \dfrac{i\omega(\mu_1-\rho(\frac{n}{2}\nu_0+\nu_1-\mu_1))+\kappa_0^2\nu_0}{M} \end{array}\right) \ .
\end{equation}

\bibliography{MassiveS1Schw}
\bibliographystyle{JHEP}

\end{document}